# Honeycomb lattice in metal-rich chalcogenide Fe$_2$Te

Jia-Qi Guan (关佳其)[1], Li Wang (王利)[2], Pengdong Wang (王鹏栋)[2], Wei Ren (任伟)[2], Shuai Lu (卢帅)[2], Rong Huang (黄荣)[2], Fangsen Li (李坊森)[2, †], Can-Li Song (宋灿立)[1,3, †], Xu-Cun Ma (马旭村)[1,3], Qi-Kun Xue (薛其坤)[1,3,4,5]

[1] State Key Laboratory of Low-Dimensional Quantum Physics, Department of Physics, Tsinghua University, Beijing 100084, China

[2] Vacuum Interconnected Nanotech Workstation (Nano-X), Suzhou Institute of Nano-Tech and Nano-Bionics (SINANO), Chinese Academy of Sciences (CAS), Suzhou 215123, China

[3] Frontier science center for quantum information, Beijing 100084, China

[4] Beijing academy of quantum information sciences, Beijing 100193, China

[5] Southern university of science and technology, Shenzhen 518055, China

†Corresponding authors: E-mail: fsli2015@sinano.ac.cn (F. Li); clsong07@mail.tsinghua.edu.cn (C-L. Song)

## Abstract

Two-dimensional honeycomb crystals have inspired intense research interest for their novel properties and great potential in electronics and optoelectronics. Here, through molecular beam epitaxy on SrTiO$_3$(001), we report successful epitaxial growth of metal-rich chalcogenide Fe$_2$Te, a honeycomb-structured film that has no direct bulk analogue, under Te-limited growth conditions. The structural morphology and electronic properties of Fe$_2$Te are explored with scanning tunneling microscopy and angle resolved photoemission spectroscopy, which reveal electronic bands cross the Fermi level and nearly-flat bands. Moreover, we find a weak interfacial interaction between Fe$_2$Te and the underlying substrates, paving a newly developed alternative avenue for honeycomb-based electronic devices.

# Introduction

Following the advent of graphene[1], two-dimensional (2D) crystals with a honeycomb lattice have been at the forefront of materials of physics, thanks to their novel and potentially useful electronic structure[2] . Specifically, the unique honeycomb lattice leads to Dirac-like bands where the charge carriers behave as massless particles [3]. As 2D analogs of graphene, a few monoelemental Xenes, to wit, borophene[4] , silicene[5], germanene[6] , stanene[7,8], phosphorene[9] , arsenene[10], antimonene[11], bismuthene[12], and tellurene[13], have been constantly synthesized and attracted worldwide attention for their unusual quantum spin hall effect, superconductivity and thermoelectric properties[14–16]. Unlike graphene, however, most of these Xenes can only be stabilized as an adhesion layer on some specific substrates, seldom form a van der Waals layered bulk phase and readily develop as freestanding 2D sheets. Alternatively, honeycomb lattice compounds of two or more elements have been becoming increasingly popular as promising candidates for tunable Dirac cones. They, for example, include transition metal chalcogenides (TMCs) that commonly crystallize into the layered structure [17,18].

As a representative TMC material, iron telluride exhibits multiple structural phases with distinct properties [19]. For instance, the conventional phase of $\beta$-FeTe with a tetragonal PbO-type structure is an antiferromagnetic metal below around 70 K [20,21], whereas the hexagonal $\alpha$-FeTe phase is ferromagnetic with the NiAs-type crystal structure [22]. In addition, iron ditelluride $FeTe_2$ is typically formed in nonlayered marcasite phase with an orthorhombic structure at ambient condition and considered as an antiferromagnetic semiconductor [23,24], although it becomes ferromagnetic at the 2D limit [25]. As the electronic and magnetic properties of Fe-Te binary compounds are hypersensitive to minor change in stoichiometry and structure, phase-controlled preparation of them have been recently explored [26–28], primarily confined to the bulk stable phases. In this Letter, we have employed Te-limited growth conditions to epitaxially prepare a metal-rich telluride $Fe_2Te$ on $SrTiO_3(001)$ substrates. *In situ* scanning tunneling microscopy (STM) and angle-resolved photoemission spectroscopy (ARPES) investigations reveal a honeycomb lattice of $Fe_2Te$, featured by electron-like bands cross the Fermi level ($E_F$) and extremely flat bands, which has no direct analogue to its bulk phase.

**Experimental Method**

Our experiments were conducted on a commercial ultrahigh vacuum (UHV) STM apparatus (Unisoku), connected to a molecular beam epitaxy (MBE) chamber for in-situ sample growth. The base pressure is better than $1.0 \times 10^{-10}$ Torr. The Nb-doped $SrTiO_3$(001) substrates were firstly degassed at 600oC, and subsequently annealed at 1250°C under UHV for 20 minutes to get the clean and flat surface. Epitaxial films of metal-rich $Fe_2Te$ were obtained by co-evaporating high-purity Fe (99.9999%) and Te (99.9999%) from standard Knudsen cells. All STM measurements were carried out at 4.7 K using sharp polycrystalline PtIr tips, which were firstly cleaned by electron beam bombardment in MBE and appropriately calibrated on Ag/Si(111) films. All STM topographic images were acquired in a constant-current mode with the bias $V$ applied to the sample. Tunneling spectra were measured using a standard lock-in technique with a small bias modulation of 10 meV at 919 Hz. After the STM measurements, the samples were transferred to vacuum-interconnected X-ray photoelectron spectroscopy (XPS) and angle-resolved photoemission spectroscopy (ARPES) chambers for further analysis of their composition and electronic structure. The ARPES measurements were performed at 78 K with a photo energy of 21.2 eV, while elemental sensitive XPS measurements in a PHI5000 system equipped with a monochromatic Al-K$a$ radiation of 1486.7 eV. All XPS spectra were calibrated using the C 1s peak at 284.6 eV and analyzed via a CasaXPS software.

**Results and Discussion**

We started with the epitaxial growth of FeTe thin films under Te-rich growth condition, with the temperatures of Fe-cell ($T_{Fe}$) and Te-cell ($T_{Te}$) set to 1040°C and 230°C, respectively. This leads to a Te/Fe flux ratio of about 10. With increasing substrate temperature $T_{sub}$, a structural transition from tetragonal $\beta$-FeTe to hexagonal $\alpha$-FeTe phases is revealed around 300°C. This differs from its sister compound FeSe [29], for which the tetragonal phase is more energetically favorable at elevated temperatures. Nevertheless, the extra Te cannot be incorporated into the 1:1 stoichiometric FeTe film and the MBE growth rate is limited by the $T_{Fe}$ and therefore Fe flux, due to the higher $T_{sub}$ than $T_{Te}$ used [27].

As the $T_{Te}$ is significantly lowered to 200°C and the MBE epitaxial growth becomes to be Te-limited (i.e. the Te/Fe flux ratio is lower than unity), we surprisingly discover a completely new phase of Fe-Te binary compound with a honeycomb lattice, as illustrated in Figs. 1(a) and 1(b). Note that the honeycomb-structured Fe-Te is not limited to monolayer films and can be stabilized to multilayers. By optimizing the $T_{sub}$ ranging from 280°C to 370°C, the metal-rich Fe-Te films almost cover the whole surface of SrTiO$_3$ substrate [Fig. 1(a)], and its out-of-plane lattice parameter is measured to be approximately 3.5 Å by taking the line profile from the honeycomb-structured Fe-Te films to underlying SrTiO$_3$ substrate [Fig. 1(c)]. High-resolution STM topographies reveal a sample bias-independent honeycomb structure [Fig. 2] with an in-plane lattice parameter of 11.5 ± 0.4 Å [Fig. 1(d)]. It is also worth noting that two distinct domains are invariably identified and are related by a 30° rotation with each other [Fig. 1(e)]. The spatial periodicity of honeycomb lattice and 30° angle of the mutual domain rotation are further corroborated by calculating the combined Fast-Fourier transform (FFT) power spectrum in Fig. 1(f), derived from the STM image in Fig. 1(e), with colored arrows distinguishing the rotational domains. Given the SrTiO$_3$(001) substrates used, we here argue that the different domains arise from symmetry mismatch between the honeycomb-structured Fe-Te thin films and square lattice substrate.

We then performed *in-situ* elemental-sensitive XPS experiments to precisely determine the stoichiometry of epitaxial honeycomb films. For comparison, similar XPS spectra have been collected on the *β*-FeTe epitaxial film as well. Figure 3 shows the two sets of XPS spectra in the Fe-2p and Te-3d core level regions. The binding energies are routinely extracted from pseudo-Voigt function fit to the spectra (see the solid lines in Fig. 3), following the subtraction of a Shirley background. As has been documented previously, the Fe-2p$_{3/2}$ and Fe-2p$_{1/2}$ peaks of *β*-FeTe located at the binding energies of 707.6 eV and 720.7 eV arise from Fe$^0$, while the Te-3d$_{5/2}$ and Te-3d$_{3/2}$ peaks located at 572.8 eV and 583.3 eV arise from Te$^0$ [30,31]. Compared with *β*-FeTe, we note negligibly small shifts (0.1 ~ 0.3 eV) of both Fe-2p and Te-3d peaks to the lower energies for the honeycomb films. This interestingly demonstrates that both Fe and Te are also zero valences at the surface of the honeycomb Fe-Te films. More remarkably, the Te-3d XPS peaks of the honeycomb films are apparently weaker than those of *β*-FeTe [Fig. 3(b)], although only a tiny difference occurs for the Fe-2p XPS peaks in the two compounds [Fig. 3(b)]. This hints at a stoichiometry of Fe:Te higher than unity in the honeycomb films. Such result is not surprising when considering the fact that the formation of honeycomb films is promoted by a substantial lowering of the Te flux rate. A quantitative analysis by measuring the peak areas and by applying appropriate atomic sensitivity factors gives rise to an Fe/Te ratio of 0.96 in $\beta$-FeTe, as reasonably anticipated, whereas the same method yields an Fe/Te ratio of 1.97 in the honeycomb films. This leads us to conclude that the epitaxial honeycomb films have a composition stoichiometry of $Fe_2Te$ that exhibits no direct bulk analogue [19]. This attribute is also consistent with the post-growth annealing experiment. Once if we annealed the epitaxial $Fe_2Te$ films under Te atmospheres at 280°C, they were all practically transformed into the tetragonal $\beta$-FeTe phase.

Having identified the chemical composition of the honeycomb films as metal-rich $Fe_2Te$ (abbreviated hereafter as $h$-$Fe_2Te$), we study its electronic structure by combining real space STM and reciprocal space ARPES techniques. Irrespective of the variation of film thickness, the $h$-$Fe_2Te$ samples are all characteristic of finite or metallic-like electronic density of states (DOS) near $E_F$, as confirmed by *in-situ* scanning tunneling spectroscopy in Fig. 4(a). The electronic states appear to be more prominent at higher energies, prompting two DOS humps at -0.50 eV and at 0.4 eV (marked by the down arrows). In addition, a dip of electronic DOS emerges around $E_F$ and is possibly caused by electron localization. In order to shed light on the metallic states and DOS humps, we further employed ARPES to explore momentum-resolved electronic structure along the K-Γ-K direction in the Brillouin zone of $Fe_2Te$ at 78 K [Fig. 4(b)], as illustrated in Figs. 4(c) and 4(d). An electron-like band dispersion (marked by the red dashes in Fig. 4(c) is revealed to pass through $E_F$ and responsible for the $E_F$ near electronic DOS observed above. The bands are more clearly seen from the differential ARPES intensity map in Fig. 4(d), from which we readily deduce the band bottom at ∼0.45 eV below $E_F$. Furthermore, an extremely flat band, which often develops in honeycomb electron lattice [32–34], is found to be located at -0.50 eV. This coincides energetically very well with the DOS hump below $E_F$ in Fig. 4(a). We therefore ascribe the DOS hump at -0.5 eV to the occurrence of nearly-flat bands or van Hove singularity in $h$-$Fe_2Te$. Further theoretical calculation is needed to fully understand the electronic structure of the honeycomb $Fe_2Te$ phase.

Provided that the in-plane lattice constant (∼11.5 Å) of $h$-$Fe_2Te$ happens to be three times as large as that (∼3.82 Å) of $\alpha$-FeTe, one might argue a fragility of $\alpha$-FeTe induced by surface reconstruction for the metal-rich $h$-$Fe_2Te$. However, this is not true, because a simultaneous imaging of the α-FeTe and h-Fe$_2$Te lattices reveals that the crystal directions of h-Fe$_2$Te run at approximately 30° to, rather than along those of α-FeTe [Fig. 4(a)]. As thus, the incommensurate epitaxial growth of h-Fe$_2$Te on the tetragonal SrTiO$_3$ substrates is intriguing and commonly expected for a weak interfacial interaction, which could effectively lift the restriction of rigid lattice match and even lattice symmetry. It is further revealed that the weak interfacial interaction seems generic to the epitaxial h-Fe$_2$Te films. When we firstly prepared β-FeTe films on the SrTiO$_3$ substrates under Te-rich growth conditions and h-Fe$_2$Te under Te-limited growth conditions at 300°C, we found similar results in Fig. 5(b). Two distinct h-Fe$_2$Te domains develop on β-FeTe and are perpendicular to each other.

## Conclusions

In summary, we have used MBE to successfully prepare metal-rich Fe$_2$Te thin films with honeycomb lattice under Te-limited growth conditions. *In-situ* STM and ARPES measurements combine to reveal electron-like bands cross $E_F$ and nearly-flat bands below $E_F$. An inspection of the epitaxial relationship demonstrates a weak interfacial interaction between the epitaxial Fe$_2$Te overlayer and underlying substrates. This finding, as well as the sizable unit cell of honeycomb lattice, might serve a novel platform for exploring the honeycomb lattice-based electronic applications, although further investigation is needed to pin down the crystal structure of honeycomb Fe$_2$Te.

## Acknowledgements

We thank J. H. Fu and Y. Xu for fruitful discussions. The work is financially supported by the National Natural Science Foundation of China (Grants No. 51788104, No. 11604366, No. 11774192, No. 11634007) and the Ministry of Science and Technology of China (Grants No. 2017YFA0304600, No. 2018YFA0305603).

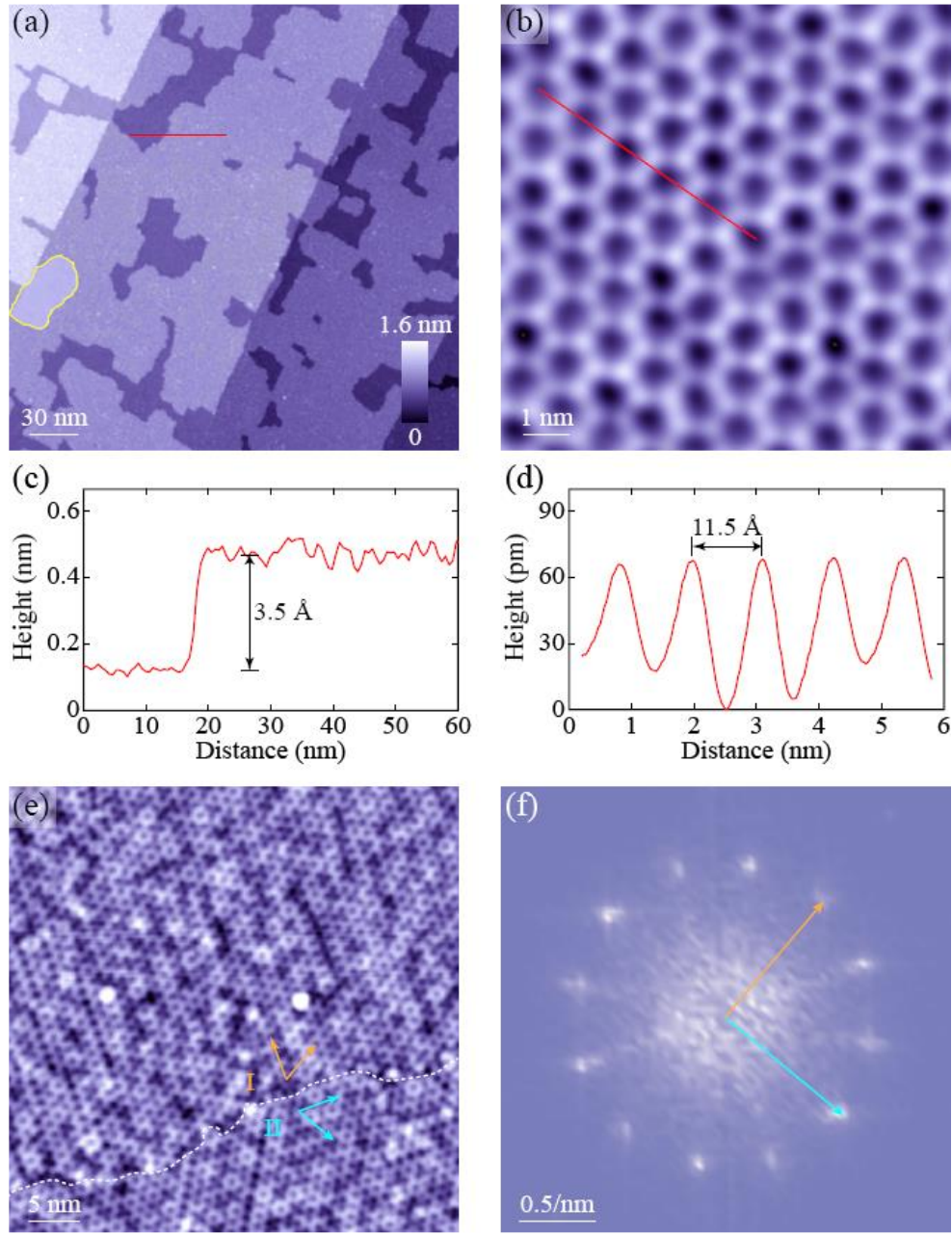

**Fig. 1**. (color online) (a) Representative STM topography (280 nm × 280 nm, V = 3.0 V, I = 50 pA) of epitaxial honeycomb-structured $Fe_2Te$ films at 330°C under Te-limited growth conditions. A small patch of hexagonal α-FeTe phase is encircled in yellow. (b) High-resolution STM image (10 nm × 10 nm, V = 0.3 V, I = 100 pA) displaying the honeycomb lattice of $Fe_2Te$. (c,d) Line profiles taken along the red lines in (a) and (b), respectively, from which the out-of-plane and in-plane lattice parameters can be readily extracted. (e) STM topography (45 nm × 45 nm, V = 2.0 V, I = 100 pA) of $Fe_2Te$ films showing two distinct domains (I and II), which are rotated by 30°C with each other (see the colored arrows). (f) Fourier transform image of STM topography in (e). The orange and cyan arrows with different FFT intensities distinguish the two rotational domains in (e).

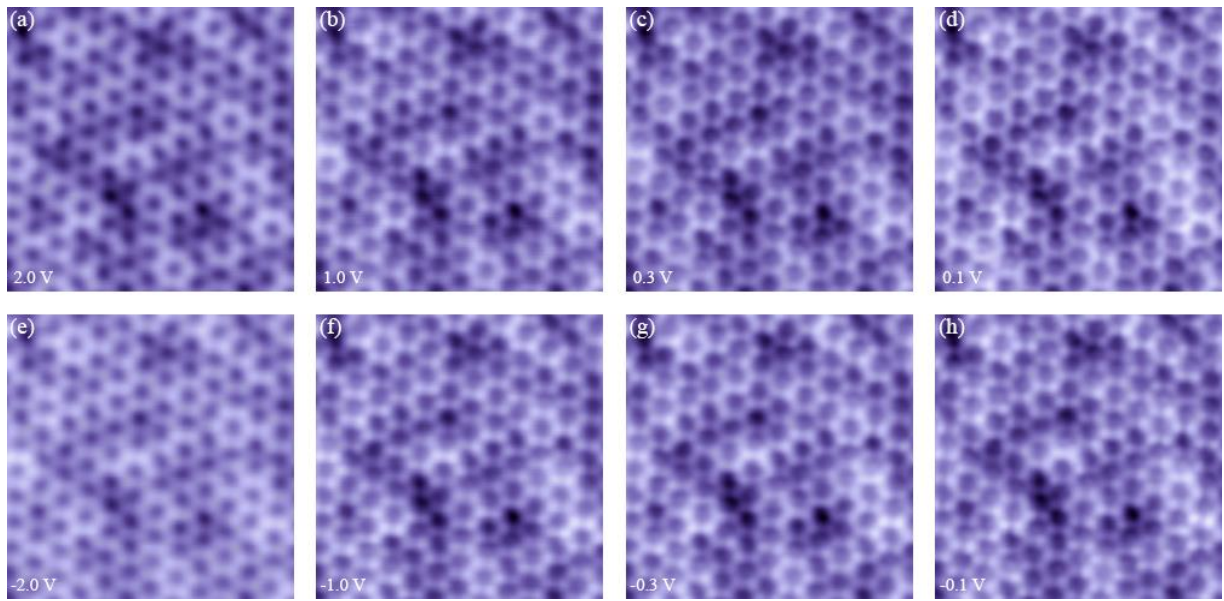

**Fig. 2.** (color online) (a-h) Atomically-resolved STM topographies (12 nm × 12 nm, $I$ = 100 pA) of $h$-Fe$_2$Te under various sample biases as indicated.

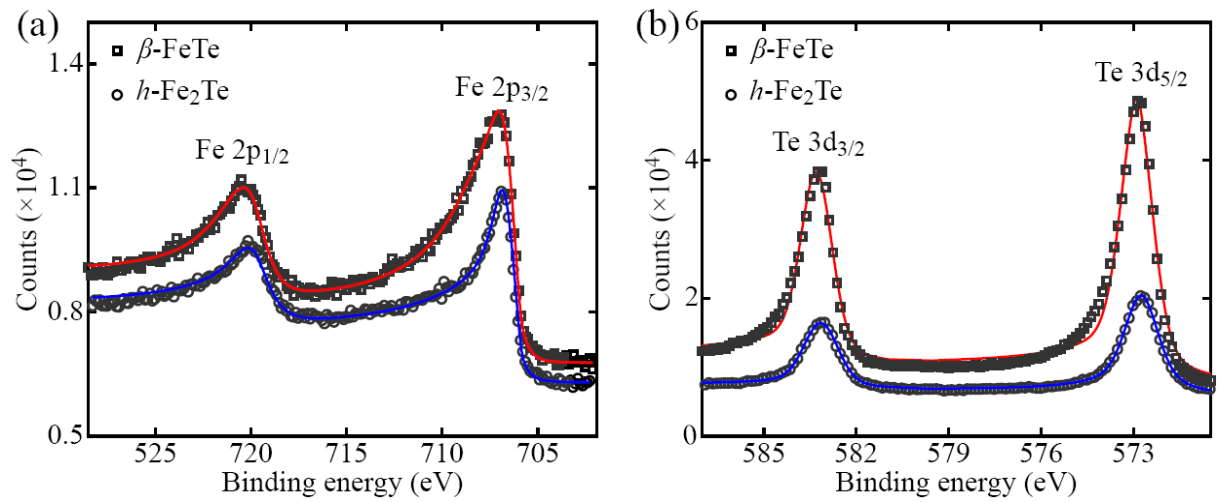

**Fig. 3**. (color online) (a, b) *In-situ* XPS spectra of tetragonal *β*-FeTe and honeycomb $Fe_2Te$ in the Fe-2p and Te-3d core level regions, respectively. Asymmetrical Voigt functions, denoted by the solid lines, nicely fits the experimental data (empty squares and circles).

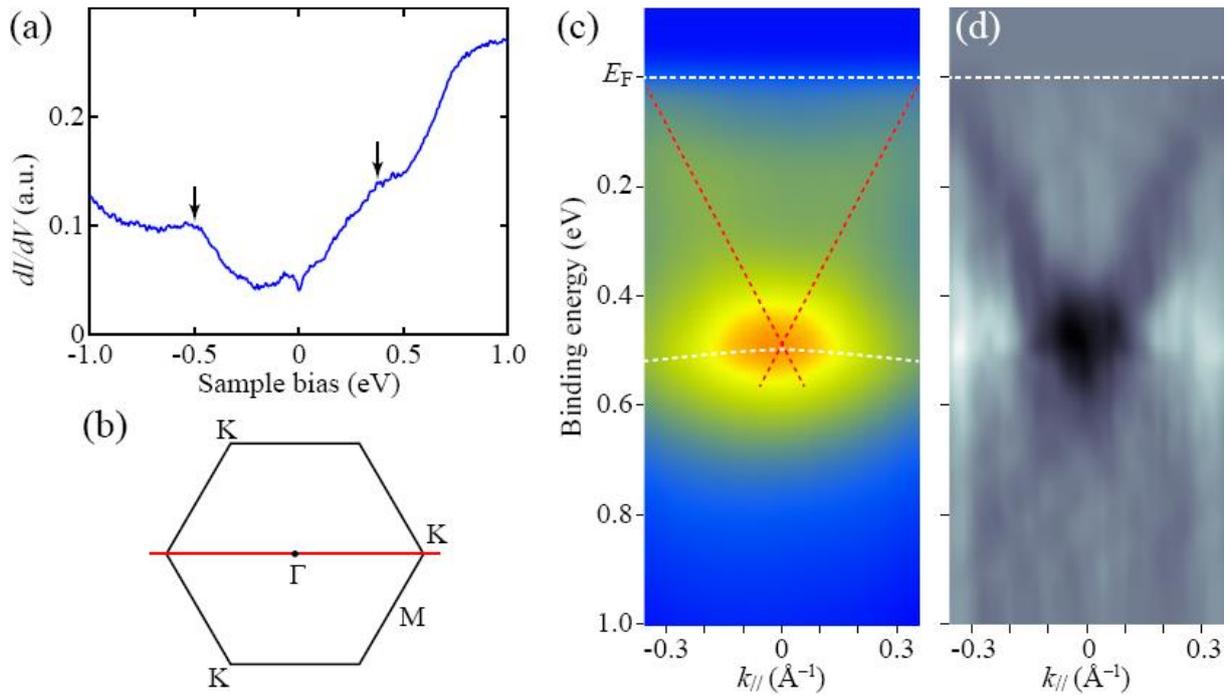

**Fig. 4**. (color online) (a) Spatially-averaged differential conductance $dI/dV$ spectrum on $h$-Fe$_2$Te, showing electron-like electronic states around $E_F$. The down arrows mark the electronic DOS humps at -0.5 eV and 0.4 eV. Tunneling junction is stabilized at $V = 1.0$ V and $I = 100$ pA. (b) Illustration of the Brillouin zone of the honeycomb lattice. (c) ARPES intensity map and (d) its differential image of Fe$_2$Te films taken along the K-Γ-K direction, as marked by the red solid line in (b). The red dashes schematically denote the band dispersion near the point, while the white one marks an extremely flat band around -0.5 eV.

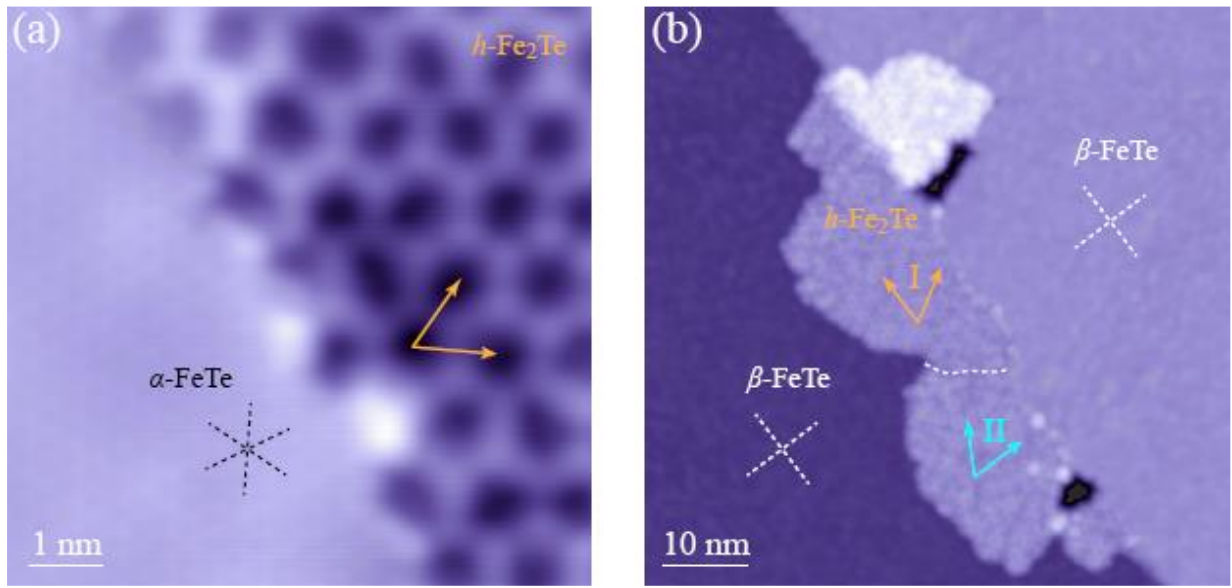

**Fig. 5**. (color online) (a) Atomically-resolved STM topography (8 nm × 8 nm, $V$ = 0.3 V, $I$ = 200 pA) across a phase boundary between the co-existing $h$-Fe2Te and $a$-FeTe films. (b) STM Topography (70 nm × 70 nm, $V$ = 1.0 V, $I$ = 100 pA) of $h$-Fe2Te grown on a pre-existing $β$- FeTe film supported by the SrTiO3 substrate. The colored arrows, black and white dashes are respectively aligned along the crystal directions of $h$-Fe2Te, $a$-FeTe and $β$-FeTe.